\documentclass[sigconf]{acmart}
\usepackage{todonotes}
\usepackage{lipsum}
\usepackage{color, colortbl}
\usepackage{multirow}
\usepackage{hyperref}

\usepackage{graphicx}
\usepackage{caption}
\usepackage{subcaption}
\usepackage{float}
\usepackage{enumitem}

\newcommand{\partitle}[1]{\vspace{2mm}\noindent\textbf{#1}}

\AtBeginDocument{%
  \providecommand\BibTeX{{%
    \normalfont B\kern-0.5em{\scshape i\kern-0.25em b}\kern-0.8em\TeX}}}

\copyrightyear{2020} 
\acmYear{2020} 
\setcopyright{acmlicensed}\acmConference[ICTIR '20]{Proceedings of the 2020 ACM SIGIR International Conference on the Theory of Information Retrieval}{September 14--17, 2020}{Virtual Event, Norway}
\acmBooktitle{Proceedings of the 2020 ACM SIGIR International Conference on the Theory of Information Retrieval (ICTIR '20), September 14--17, 2020, Virtual Event, Norway}
\acmPrice{15.00}
\acmDOI{10.1145/3409256.3409817}
\acmISBN{978-1-4503-8067-6/20/09}

\begin{document}

\fancyhead{}
\fancyfoot{}

\title{Analysing~the~Effect~of~Clarifying ~Questions on~Document~Ranking~in~ Conversational~Search}

\newcommand{\AuthorSpace}{ \ \ \ \ \ }

\author{%
Antonios~Minas~Krasakis\AuthorSpace
Mohammad~Aliannejadi\AuthorSpace
Nikos~Voskarides\AuthorSpace
Evangelos~Kanoulas
}
\affiliation{%
  \institution{
  University of Amsterdam
  }
}  
\email{{a.m.krasakis, m.aliannejadi, n.voskarides, e.kanoulas}@uva.nl}

\def\authors{Antonios~Minas~Krasakis, Mohammad~Aliannejadi, Nikos Voskarides, Evangelos Kanoulas}




\renewcommand{\shortauthors}{Krasakis, et al.}


\begin{abstract}

Recent research on conversational search highlights the importance of mixed-initiative in conversations. 
To enable mixed-initiative, the system should be able to ask clarifying questions to the user. 
However, the ability of the underlying ranking models (which support conversational search) to account for these clarifying questions and answers has not been analysed when ranking documents, at large. To this end, we analyse the performance of a lexical ranking model on a conversational search dataset with clarifying questions.
We investigate, both quantitatively and qualitatively, how different aspects of clarifying questions and user answers affect the quality of ranking.
We argue that there needs to be some fine-grained treatment of the entire conversational round of clarification, based on the explicit feedback which is present in such mixed-initiative settings. Informed by our findings, we introduce a simple heuristic-based lexical baseline, that significantly outperforms the existing naive baselines.
Our work aims to enhance our understanding of the challenges present in this particular task and inform the design of more appropriate conversational ranking models.
\end{abstract}


\maketitle

\section{Introduction}


The rise of voice-based digital assistants such as Amazon Alexa and Google Assistant, has intensified the need for agents that can hold meaningful conversations with users. Towards this direction, researchers have developed conversational systems that support question-answering and task-oriented dialogue, among others~\cite{budzianowski2018multiwoz,DBLP:conf/emnlp/ChoiHIYYCLZ18}. However, it is often the case that in such information-seeking conversations, users fail to express their information need adequately.
This makes the ability of a conversational search system to support mixed-initiative interactions imperative~\cite{radlinski2017theoretical, DBLP:conf/sigir/KieselBSAH18}.
%
Such a system can assist users to refine their information need, i.e., by disclosing new information to them~\cite{radlinski2017theoretical}, or posing clarifying questions~\cite{aliannejadi2019asking}.

\begin{figure}[t]
    \centering
    \vspace{-7mm}
    \includegraphics[scale=0.35]{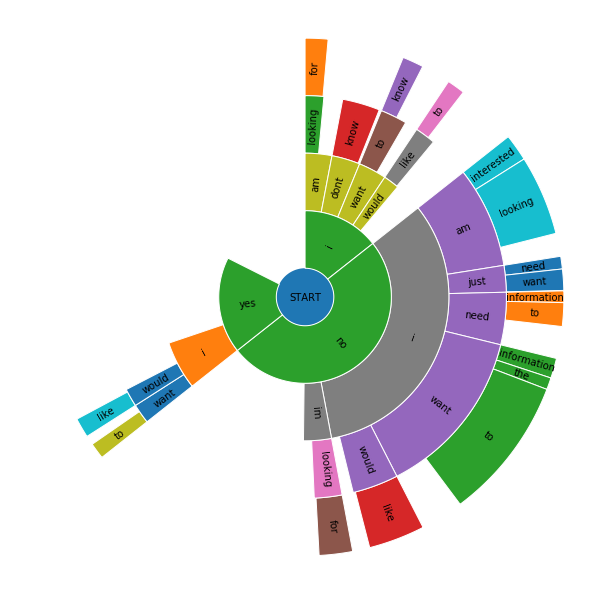}
    \vspace{-8mm}
    \caption{Most frequent 4-grams to answers of clarifying questions. Each word covers an arc proportional to its frequency.}
    \label{fig:4grams}
    \vspace{-6mm}
\end{figure}

Clarifying questions trigger users' explicit feedback in the form of an answer, and have been shown to improve user experience ~\cite{aliannejadi2019asking,DBLP:conf/sigir/KieselBSAH18, zamani2020analyzing,DBLP:conf/chiir/BraslavskiSAD17}. 
In Figure~\ref{fig:4grams}, we demonstrate examples of clarification-based conversations appearing in the conversational search dataset Qulac~\cite{aliannejadi2019asking}. Specifically, we plot the most frequent user responses ($4-grams$) to clarifying questions. 
Responses can be read starting from the circle centre ({\small{$START$}}) and moving outwards (e.g., ``no I am looking...''). We observe that user responses often start with a ``yes'' or a ``no'', but frequently provide additional information (e.g., ``No, I want...'').
We hypothesise that this explicit feedback can be used to improve ranking, even when the question asked ranges from being partially relevant to completely irrelevant w.r.t. the information need. %

However, using this feedback effectively is challenging due to: (i) the noisy nature of the natural language used in mixed-initiative conversations, (ii) the presence of complex and mixed (both positive \& negative) signals that answers often convey (e.g., user: ``Yes, but I would like ...''), and
(iii) the presence of partially relevant information, in clarification questions that have received negative user feedback (e.g., {\small{$[inf. need]$}}: "Drafting of Declaration of Independence",  {\small{$[system]$}}: ``Would you like to learn more about Thomas Jefferson?'' {\small{$[user]$}}: ``No.'').

In this paper, we study the effect of the user's feedback in mixed-initiative conversations.
We categorise answers w.r.t. their polarity and length. Answer polarity indicates whether the question points to a relevant direction or not, while answer length enables us to (noisily) identify the presence of additional information in the response. We conduct our analysis on the Qulac dataset~\cite{aliannejadi2019asking}, using a query likelihood model chosen because of its simplicity and transparency~\cite{ponte1998language}.

In particular, we aim to answer the following research questions:
{\bf RQ1} How does the polarity and informativeness of the user's response to a system's clarifying question affect the performance of a term-matching ranking model in conversational search?
{\bf RQ2} How does the length of the clarifying question or the user's response affect the performance of a term-matching ranking model in conversational search?
{\bf RQ3} How well can a simple rule-based ranking model perform, compared to existing baselines?



Our contributions are as follows:
(i) we study the answers that users provide in a conversation, and categorise them based on polarity and length using simple yet effective heuristics;
(ii) we conduct an in-depth analysis of the performance of the QL model, for those answer types;
(iii) we study the effect of the clarification question and answer on ranking performance, when used in isolation, combined or ignored, and
(iv) we design a simple heuristic ranker, which outperforms the baseline lexical model significantly, solely by incorporating information about the answer type.


\section{Experimental Setup}

\partitle{Data.}
We run our experiments on the task of document ranking. Different from typical ad-hoc search, here we assume that after the initial query ($Q_0$) posed by a user, a clarifying question ($Q$) is asked by the system, followed by a user's answer ($A$). We use the document corpus and single-round conversations provided by Qulac, the only large-scale conversational search dataset with clarifying questions we are aware of~\cite{aliannejadi2019asking}. 
Qulac is built on top of the TREC Web Track 2009-2012 data and consists of 10K query-question-answer tuples for 198 TREC topics with 762 facets, with each Q\&A corresponding to a facet. We randomly keep 40 topics for testing the performance of the heuristic ranker (section~\ref{sec:heuristic-ranker}) and perform all our analysis on the rest of the topics.



\partitle{Retrieval Model.}
We use a KL divergence query-likelihood model with Dirichlet prior smoothing~\cite{ponte1998language}. 
To adapt this model in a conversational setting, we initially experimented with expanding $Q_0$ with the rest of the clarification round ($Q$ and $A$). However, preliminary results indicated poor ranking performance, since the importance of the topic (expressed by $Q_0$) is largely underestimated.
To mitigate this issue, we follow \cite{aliannejadi2019asking} and linearly interpolate the original query ($Q_0$) with the clarification round ($Q$ and $A$, concatenated) using an equal score interpolation weight of $0.5$.
We use $NDCG@20$ for evaluation.

\section{Results}

In this section, we present our experimental results w.r.t. the RQs.

\vspace{-2mm}
\subsection{Answer polarity and informativeness} 




In this section we aim to answer {\bf RQ1} by measuring the  performance of the retrieval model described in the previous section per answer type, using different parts of the conversation round.
To do so, we identify two important characteristics of the user's response to a system's clarifying question, namely: (i) the polarity of the answer, and (ii) the informativeness of the answer. 
Answer polarity gives us a strong signal about the question being asked, and whether it was relevant to the user's information need.
We define four polarity classes:  Positive $P$, Negative $(N)$, ``I don't know'' $(idk)$, and Other $(O)$.\footnote{The \textit{Other} category refers to the answers that did not fit the other three categories.}
Motivated by Figure~\ref{fig:4grams}, we use a simple yet effective heuristic to annotate positive and negative polarity in an answer. More specifically, we tag an answer with $P$ if it contains the term ``yes'', and with $N$ if it contains the term ``no''.

\begin{table*}[t]
    \centering
    \caption{QL performance per answer type and length (NDCG@20). $Q_0$ is the original query, $Q$ is the clarification question and $A$ is the answer to the clarifying question. Single refers to answers of length 1 and multi refers to answers of length greater than 1. We measure $\Delta$  w.r.t.  $Q_0$ and indicate significant differences (2-sided $t-test$, $p-value < 0.05$) with \dag. }
\label{tab:results-polarity-length}
\vspace{-3mm}
\begin{tabular}{rrr||l|ll|ll|ll}
\toprule
Answer Polarity    &  Answer Length   &     \# samples &        $Q_0$ &        $Q_0 + Q$ &  &     $Q_0 + A$   &  &  $Q_0 + Q + A$ &  \\
    &     &      & &         & $\Delta$ (\%) &        & $\Delta$ (\%) &   & $\Delta$ (\%) \\
\midrule

$P$   & single               & 364  & 0.191          & \textbf{0.206} & +7.9$^\dag$  & 0.191          & +0.0                         & \textbf{0.206} & +7.9$^\dag$  \\
    & multi & 1275 & 0.162          & 0.162          & +0.0                         & 0.163          & +0.6                         & \textbf{0.188} & +16.0$^\dag$ \\ \hline
$N$   & single               & 580  & \textbf{0.130} & 0.106          & $-$18.5$^\dag$ & 0.129          & $-$0.8                         & 0.106          & $-$18.5$^\dag$ \\
    & multi & 3791 & 0.132          & 0.117          & $-$11.4$^\dag$ & \textbf{0.162} & +22.7$^\dag$ & 0.159          & +20.5$^\dag$ \\ \hline
$O$   & single               & 47   & \textbf{0.177} & 0.147          & $-$16.9                        & 0.085          & $-$52.0$^\dag$ & 0.170          & $-$4.0                         \\
    & multi & 1729 & 0.153          & 0.132          & $-$13.7$^\dag$ & 0.169          & +10.5$^\dag$ & \textbf{0.171} & +11.8$^\dag$ \\ \hline
$idk$ & multi & 346  & \textbf{0.162} & 0.141          & $-$13.0$^\dag$ & 0.109          & $-$32.7$^\dag$ & 0.143          & $-$11.7$^\dag$                       \\\bottomrule

\end{tabular}
\end{table*}


\subsubsection{Ranking with full conversation.}
Here we discuss the ranker's performance when using the whole clarification round ($Q_0 + Q + A$). We discuss the results shown in Table ~\ref{tab:results-polarity-length} per answer type below, focusing on the relative improvement or decrease w.r.t. when only using $Q_0$. 

%
\partitle{Positive answers ($P$).} We observe that performance is improved when the answer is positive  for both single-word and multi-word answers.
This suggests that both the question and answer contain terms that complement the description of the information need and help the QL model rank relevant documents higher.
\partitle{Negative answers ($N$).}
We observe that the performance drops substantially when the user answer is single-word (``no''). In contrast, the user answer is multi-word, the clarification round significantly improves the ranking performance. This indicates that even in the presence of conflicting or misleading clarifying questions, a simple term-matching model can benefit from the user's answer.
\partitle{Other answers ($O$).}
We observe an improvement when the user answer is multi-word, similarly to what we observed in the negative answers.
\partitle{``I don't know'' answers ($idk$).}
We observe that it is preferable to ignore the clarification round altogether. This is likely related to the data collection strategy followed in \cite{aliannejadi2019asking}, where the crowd-workers were instructed to respond ``I don't know'' to questions that cannot be answered in the context of the annotation task. Such questions include personal questions (e.g., ``Do you have diabetes?'') or questions irrelevant to the information need.
%

\subsubsection{Ranking with part of the conversation.}

In this experiment, we aim to investigate how the performance of the QL ranker is affected when we take into account the clarification round \textit{partially}, i.e., when we ignore the answer ($Q_0+Q$) or the question ($Q_0+A$). To this end, we first study how the performance is affected when we have positive multi-word answers, followed by cases where the answer polarity changes.

\begin{figure}[t]
	\centering
	\includegraphics[width=0.8\columnwidth]{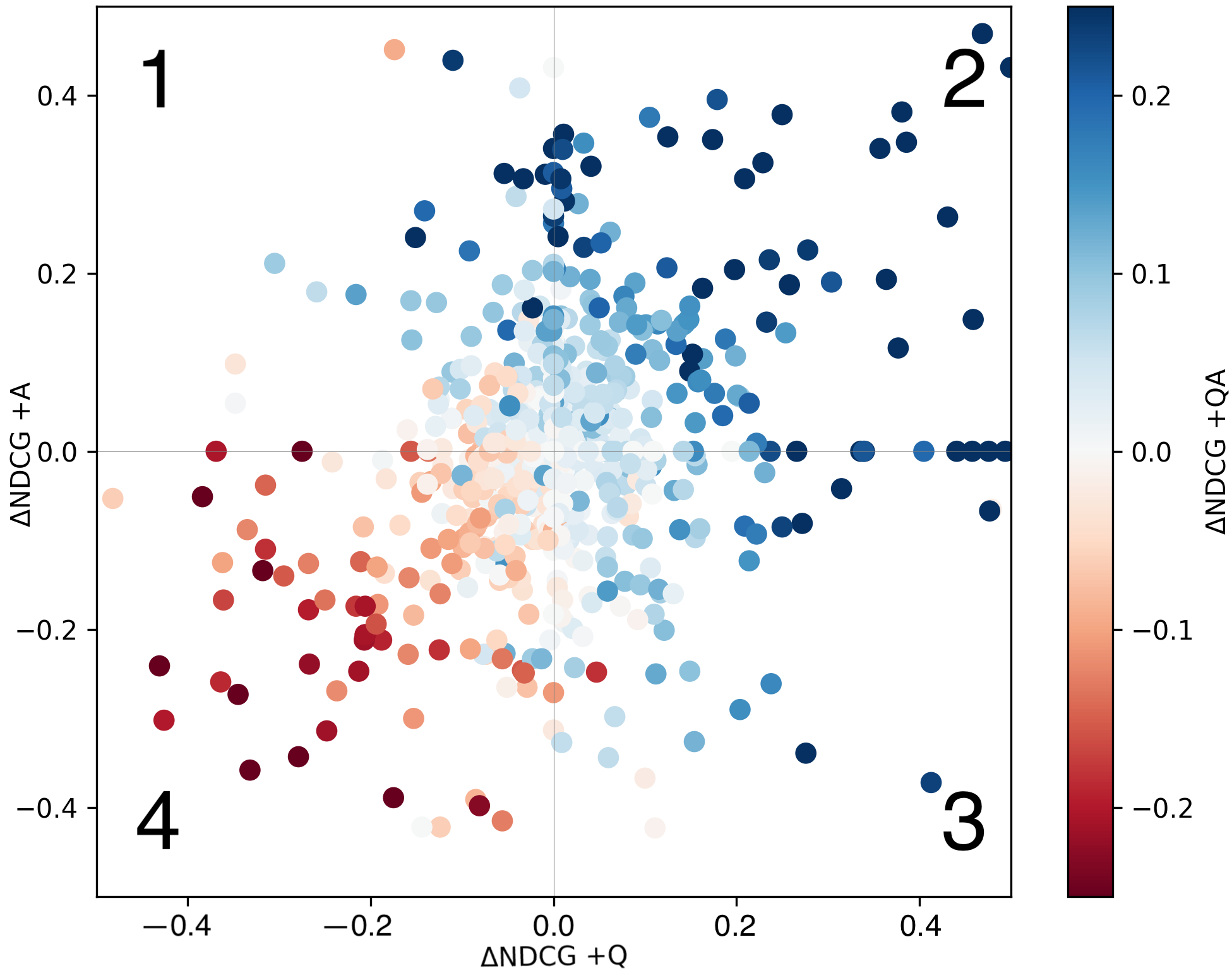}
	\caption{Scatter plot of $\Delta NDCG@20 $ (w.r.t. $Q_0$) on long, positive answers.}
	\label{fig:scatter-positive-long}
\end{figure}

\partitle{Positive multi-word answers.} In Table~\ref{tab:results-polarity-length}, we see that ignoring part of the clarification round ($Q_0+Q$ or $Q_0+A$) seems to diminish any improvements that would occur in the presence of the entire round ($Q_0+Q+A$). This indicates that in those cases the value of the clarification does not exist in isolation in questions or answers, but in their combination, implying that the question and the answer could contain complementary information. 
To further investigate this, in Figure~\ref{fig:scatter-positive-long} we plot the differences of the QL ranker when using the full clarification round ($\Delta NDCG (+Q+A)$) w.r.t. $Q_0$. Notice that we also plot the differences (of the same conversation) when using only the question in the x-axis ($\Delta NDCG +Q$), and only the answer in the y-axis ($\Delta NDCG +A)$. This enables us to better understand how the performance of the overall clarification is affected by the question and the answer.
We see that when $+Q$ and $+A$ follow the same trend -- either improve (see quarter 2) or harm the performance (see quarter 4), their combination follows with a few exceptions.
However, when questions harm but answers improve (see quarter 1), combining them in a conversational round can either improve or harm the performance. 
In the reverse case (see quarter 3), we observe a more robust positive impact in the combination of question and answer ($+Q+A$). 
After examining a few examples, we observe that the main cause of this is that questions often contain crucial information, without which the answer is incomplete in isolation (e.g., {\small{$[Q_0]$}} pork tenderloin  {\small{$[Q]$}}``Would you like to make a \textbf{rub} for pork tenderloin?''  {\small{$[A]$}}``Yes, but I need to find a \textbf{recipe}'').

\partitle{$Q_0$ in positive vs. negative answers.} 
    Another interesting observation deriving from Table~\ref{tab:results-polarity-length} is that the performance when only using the original query ($Q_0$) is quite low for negative answers (second group) compared to positive answers (first group). It is important to highlight here that this performance is completely unaffected by the questions asked and answers received, since it only takes into account the original query.
    Our hypothesis is that this is a side-effect of the process through which the clarification questions were collected in~\cite{aliannejadi2019asking}. Specifically, the crowd-workers were instructed to read the first two result pages of a commercial search engine using the original query (topic) and compose a clarification question based on those. Therefore, this created some bias in the question generation towards more \textit{popular} information needs. Let us note here, that this bias is defined by the search engine used by each crowd-worker and could even be a desirable characteristic of the dataset.

    To validate this hypothesis, in Table~\ref{tab:correlation-q0-polarity} we analyse the percentage of positive and negative answers received per facet, w.r.t. the performance of the facet using only the original query ($Q_0$), while ignoring the clarification round ($Q$ and $A$). We see that there is a significant correlation between facet performance and the percentage of clarification questions pointed towards the correct direction. This suggests that identifying \textit{easier} facets (higher $NDCG@20$) is an easier task for the crowd-workers. This is mainly because naturally the facets with higher performance have more relevant documents at the top of the ranked list, and hence it is more likely for a crowd-worker to observe a facet in a relevant document and ask a question about it.

\begin{table}[t]
\vspace{-2mm}
\caption{Correlation between $NDCG@20$ when using $Q_0$ and percentage of positive, negative answers per facet.}
\vspace{-3mm}
\label{tab:correlation-q0-polarity}
\begin{tabular}{lrr}
\hline
     & \begin{tabular}[c]{@{}r@{}}Pearson's $r$ \end{tabular} & p-value      \\ \hline
$P$    & 0.173                                                                          & $1.5^{-6}$ \\
$N$    &                             -0.197 &  $4.3^{-8}$ \\
rest & 0.208                                                                          & $6.8^{-9}$ \\ \hline
\end{tabular}
\vspace{-4mm}
\end{table}

\subsection{Clarifying Question and Answer Length}

\begin{table}
\small
\caption{Qualitative analysis of clarification rounds with single-word negative answers.}
\label{tab:qualitative-analysis-single-negative-answers}
\vspace{-4mm}
\begin{tabular}{l|l|c}
          \multicolumn{2}{l}{}                                                    &                                          $\Delta NDCG$ \\
\toprule  
Inf. need & \textit{Drafting of Declaration of Independence}                                                    & \multirow{4}{*}{+18.53}           \\
$Q_0$     & all men are created equal                                                                       &                                   \\
$Q$        & \begin{tabular}[c]{@{}l@{}}- Would you like to learn more about \\ \textbf{Thomas Jefferson}?\end{tabular} &                                   \\
$A$    & - No                                                                                           &                                  \\\toprule
Inf. need & \textit{Information about Atari arcade games}                                                    & \multirow{4}{*}{+15.97}           \\
$Q_0$     & atari                                                                                        &                                   \\
$Q$        & \begin{tabular}[c]{@{}l@{}}- Would you like to play atari \\ \textbf{arcade games} online?\end{tabular} &                                   \\
$A$    & - No                                                                                           &                                  \\\toprule
 Inf. need & \textit{How is a credit score determined?}                                                    & \multirow{4}{*}{+13.57}           \\
 $Q_0$     & credit report                                                                                        &                    \\
 $Q$        & \begin{tabular}[c]{@{}l@{}} - Would you like to know about the process \\ of \textbf{disputing} credit report?\end{tabular} &                                   \\
 $A$    & - No                                                                                           &                                  \\\toprule
\end{tabular}
\end{table}

\begin{table}[]
\caption{Pearson correlation between the length of the question and answer w.r.t. $\Delta NDCG$ from the original query ($Q_0$).}
\label{tab:correlation-length}
\vspace{-4mm}
\begin{tabular}{ll||ll}
\toprule
      X      &        Y          & Pearson's r &  p-value \\
\midrule
\# tokens in $Q$ & $\Delta NDCG (+Q)$ &       0.071 &  $1.1^{-11}$ \\
            & $\Delta NDCG (+Q+A)$ &      0.006 &     $5.6^{-1}$ \\\hline
\# tokens in $A$  & $\Delta NDCG (+A)$ &        0.130 &  $4.4^{-34}$ \\
            & $\Delta NDCG (+Q+A)$ &       0.056 & $ 1.1^{-7}$ \\\hline
\# tokens in $Q+A$  & $\Delta NDCG (+Q+A)$ &       0.049 &  $4.1^{-6}$ \\
\bottomrule
\end{tabular}
\vspace{-4mm}
\end{table}
Here, we aim to study the effect of the length of questions and answers on the performance (\textbf{RQ2}), as it is unclear whether longer conversations would result in more informative queries and better results.
Traditional IR models have been thoroughly tested with keyword-based queries. However, in  the mixed-initiative setting many irrelevant terms would appear, even after stop-word removal.

Table \ref{tab:correlation-length} shows that significant correlations exist between the length of questions or answers and improvement in $NDCG$  when those are added to the query. 
This correlation is strongest for answers, which is expected, since users have concrete knowledge of the information need in our setup. For questions, we observe the same trend. This is important, as most answers are negative (see Table~\ref{tab:results-polarity-length}) and highlights that even questions that receive negative answers contain helpful information for ranking, since they reduce the term mismatch.
To further clarify why this happens, we provide examples where questions received negative and single-word responses (implying that no meaningful information was provided in the clarification round), but resulted in a big increase in $\Delta NDCG$ (Table~\ref{tab:qualitative-analysis-single-negative-answers}).
We observe that while asking clarifying questions, a number of contextually relevant words appear. For instance, \textbf{Thomas Jefferson} is a very relevant entity w.r.t. the drafting of the declaration of independence and their name regularly appears within the relevant results, although the query is not strictly connected to him. Likewise, \textbf{arcade games} 
are relevant words which were previously not mentioned and help improve the ranking. Even further, words such as \textbf{disputing} 
are much less relevant to the information need, but they still appear in the relevant documents because they do contain information about those too (e.g., the disputing credit reports). This indicates that taking into account such conversations has, to a certain extent, similar effects to query expansion and can prove helpful, even when the added terms are not strictly relevant to the information need.

\subsection{Heuristic Ranker}
\label{sec:heuristic-ranker}
Based on the insights from the results presented so far, we design a ranker that  heuristically classifies the user responses using polarity and length, and chooses which parts of the clarification round ($Q$, $A$) to use (\textbf{RQ3}).
Following the observations of Table \ref{tab:results-polarity-length}, the ranker interpolates the original query ($Q_o$) with:
(i) only $A$ when answers are multi-word negative $N$,
(ii) $Q + A$ for all answers that are positive, or ``Other'' ($O$) and multi-word, and (iii) ignores both question and answer (only uses $Q_o$) elsewhere, where clarifications do not seem to help. 
We evaluate our model on a held-out test set and report the results in Table~\ref{tab:heuristic-model}. %
We observe improvements compared to all of the baselines, despite the simplicity of the proposed model (the improvement w.r.t. the best compared model, $Q0+Q+A$, are significant at $p < 0.001$). 
This suggests that using more advanced techniques to understand and effectively incorporate conversational feedback is likely improve ranking in this task.

\begin{table}
\caption{Comparison of the heuristic ranker with variations of the QL model.}
\label{tab:heuristic-model}
\begin{tabular}{lc}
\toprule
Ranker &  NDCG@20  \\
\midrule
$Q0$     &    0.148  \\
$+Q$     &    0.134  \\
$+A$     &    0.163  \\
$+Q+A$    &    0.166  \\
\hline
Heuristic ranker &    {\bf 0.171}           \\
\bottomrule
\end{tabular}
\vspace{-5mm}
\end{table}

\vspace{-2mm}
\section{Conclusions and Future Work}
In this work, we provided insights on the task of document ranking with clarification-based conversations. 
We highlighted the importance of effectively understanding and incorporating explicit conversational feedback, and demonstrated challenges by quantitative and qualitative means.
We argued that a more fine-grained treatment of the conversations is crucial to the success of conversational search and propose a heuristic ranking model, which addresses part of the problem despite its simplicity.
As future work, we plan to expand our study to account for recently developed neural ranking models for this task~\cite{hashemi2020guided}. 
Also, we aim to explore more sophisticated methods for classifying answers and develop ranking models that can better incorporate explicit conversational feedback.

\begin{acks}
This research was supported by
the NWO Innovational Research Incentives Scheme Vidi (016.Vidi.189.039),
the NWO Smart Culture - Big Data / Digital Humanities (314-99-301),
the H2020-EU.3.4. - SOCIETAL CHALLENGES - Smart, Green And Integrated Transport (814961) the Google Faculty Research Awards program.
All content represents the opinion of the authors, which is not necessarily shared or endorsed by their respective employers and/or sponsors.

\end{acks}

\bibliographystyle{ACM-Reference-Format}
\bibliography{references}


\begin{thebibliography}{9}


\ifx \showCODEN    \undefined \def \showCODEN     #1{\unskip}     \fi
\ifx \showDOI      \undefined \def \showDOI       #1{#1}\fi
\ifx \showISBNx    \undefined \def \showISBNx     #1{\unskip}     \fi
\ifx \showISBNxiii \undefined \def \showISBNxiii  #1{\unskip}     \fi
\ifx \showISSN     \undefined \def \showISSN      #1{\unskip}     \fi
\ifx \showLCCN     \undefined \def \showLCCN      #1{\unskip}     \fi
\ifx \shownote     \undefined \def \shownote      #1{#1}          \fi
\ifx \showarticletitle \undefined \def \showarticletitle #1{#1}   \fi
\ifx \showURL      \undefined \def \showURL       {\relax}        \fi
\providecommand\bibfield[2]{#2}
\providecommand\bibinfo[2]{#2}
\providecommand\natexlab[1]{#1}
\providecommand\showeprint[2][]{arXiv:#2}

\bibitem[\protect\citeauthoryear{Aliannejadi, Zamani, Crestani, and
  Croft}{Aliannejadi et~al\mbox{.}}{2019}]%
        {aliannejadi2019asking}
\bibfield{author}{\bibinfo{person}{Mohammad Aliannejadi},
  \bibinfo{person}{Hamed Zamani}, \bibinfo{person}{Fabio Crestani}, {and}
  \bibinfo{person}{W~Bruce Croft}.} \bibinfo{year}{2019}\natexlab{}.
\newblock \showarticletitle{Asking clarifying questions in open-domain
  information-seeking conversations}. In \bibinfo{booktitle}{\emph{SIGIR}}.
\newblock


\bibitem[\protect\citeauthoryear{Braslavski, Savenkov, Agichtein, and
  Dubatovka}{Braslavski et~al\mbox{.}}{2017}]%
        {DBLP:conf/chiir/BraslavskiSAD17}
\bibfield{author}{\bibinfo{person}{Pavel Braslavski}, \bibinfo{person}{Denis
  Savenkov}, \bibinfo{person}{Eugene Agichtein}, {and} \bibinfo{person}{Alina
  Dubatovka}.} \bibinfo{year}{2017}\natexlab{}.
\newblock \showarticletitle{What Do You Mean Exactly? Analyzing Clarification
  Questions in {CQA}}. In \bibinfo{booktitle}{\emph{CHIIR}}.
\newblock


\bibitem[\protect\citeauthoryear{Budzianowski, Wen, Tseng, Casanueva, Ultes,
  Ramadan, and Ga{\v{s}}i{\'c}}{Budzianowski et~al\mbox{.}}{2018}]%
        {budzianowski2018multiwoz}
\bibfield{author}{\bibinfo{person}{Pawe{\l} Budzianowski},
  \bibinfo{person}{Tsung-Hsien Wen}, \bibinfo{person}{Bo-Hsiang Tseng},
  \bibinfo{person}{I{\~n}igo Casanueva}, \bibinfo{person}{Stefan Ultes},
  \bibinfo{person}{Osman Ramadan}, {and} \bibinfo{person}{Milica
  Ga{\v{s}}i{\'c}}.} \bibinfo{year}{2018}\natexlab{}.
\newblock \showarticletitle{{M}ulti{WOZ} - A Large-Scale Multi-Domain
  Wizard-of-{O}z Dataset for Task-Oriented Dialogue Modelling}. In
  \bibinfo{booktitle}{\emph{EMNLP}}.
\newblock


\bibitem[\protect\citeauthoryear{Choi, He, Iyyer, Yatskar, Yih, Choi, Liang,
  and Zettlemoyer}{Choi et~al\mbox{.}}{2018}]%
        {DBLP:conf/emnlp/ChoiHIYYCLZ18}
\bibfield{author}{\bibinfo{person}{Eunsol Choi}, \bibinfo{person}{He He},
  \bibinfo{person}{Mohit Iyyer}, \bibinfo{person}{Mark Yatskar},
  \bibinfo{person}{Wen{-}tau Yih}, \bibinfo{person}{Yejin Choi},
  \bibinfo{person}{Percy Liang}, {and} \bibinfo{person}{Luke Zettlemoyer}.}
  \bibinfo{year}{2018}\natexlab{}.
\newblock \showarticletitle{QuAC: Question Answering in Context}. In
  \bibinfo{booktitle}{\emph{EMNLP}}.
\newblock


\bibitem[\protect\citeauthoryear{Hashemi, Zamani, and Croft}{Hashemi
  et~al\mbox{.}}{2020}]%
        {hashemi2020guided}
\bibfield{author}{\bibinfo{person}{Helia Hashemi}, \bibinfo{person}{Hamed
  Zamani}, {and} \bibinfo{person}{W~Bruce Croft}.}
  \bibinfo{year}{2020}\natexlab{}.
\newblock \showarticletitle{Guided Transformer: Leveraging Multiple External
  Sources for Representation Learning in Conversational Search}. In
  \bibinfo{booktitle}{\emph{SIGIR}}.
\newblock


\bibitem[\protect\citeauthoryear{Kiesel, Bahrami, Stein, Anand, and
  Hagen}{Kiesel et~al\mbox{.}}{2018}]%
        {DBLP:conf/sigir/KieselBSAH18}
\bibfield{author}{\bibinfo{person}{Johannes Kiesel}, \bibinfo{person}{Arefeh
  Bahrami}, \bibinfo{person}{Benno Stein}, \bibinfo{person}{Avishek Anand},
  {and} \bibinfo{person}{Matthias Hagen}.} \bibinfo{year}{2018}\natexlab{}.
\newblock \showarticletitle{Toward Voice Query Clarification}. In
  \bibinfo{booktitle}{\emph{{SIGIR}}}.
\newblock


\bibitem[\protect\citeauthoryear{Ponte and Croft}{Ponte and Croft}{1998}]%
        {ponte1998language}
\bibfield{author}{\bibinfo{person}{Jay~M Ponte} {and} \bibinfo{person}{W~Bruce
  Croft}.} \bibinfo{year}{1998}\natexlab{}.
\newblock \showarticletitle{A language modeling approach to information
  retrieval}. In \bibinfo{booktitle}{\emph{SIGIR}}.
\newblock


\bibitem[\protect\citeauthoryear{Radlinski and Craswell}{Radlinski and
  Craswell}{2017}]%
        {radlinski2017theoretical}
\bibfield{author}{\bibinfo{person}{Filip Radlinski} {and} \bibinfo{person}{Nick
  Craswell}.} \bibinfo{year}{2017}\natexlab{}.
\newblock \showarticletitle{A theoretical framework for conversational search}.
  In \bibinfo{booktitle}{\emph{ICTIR}}.
\newblock


\bibitem[\protect\citeauthoryear{Zamani, Mitra, Chen, Lueck, Diaz, Bennett,
  Craswell, and Dumais}{Zamani et~al\mbox{.}}{2020}]%
        {zamani2020analyzing}
\bibfield{author}{\bibinfo{person}{Hamed Zamani}, \bibinfo{person}{Bhaskar
  Mitra}, \bibinfo{person}{Everest Chen}, \bibinfo{person}{Gord Lueck},
  \bibinfo{person}{Fernando Diaz}, \bibinfo{person}{Paul~N Bennett},
  \bibinfo{person}{Nick Craswell}, {and} \bibinfo{person}{Susan~T Dumais}.}
  \bibinfo{year}{2020}\natexlab{}.
\newblock \showarticletitle{Analyzing and Learning from User Interactions for
  Search Clarification}. In \bibinfo{booktitle}{\emph{SIGIR}}.
\newblock


\end{thebibliography}

\end{document}